\documentclass[traditabstract]{aa}
\usepackage{txfonts}
\usepackage{graphicx}
\usepackage{natbib}
\usepackage{epstopdf}

\AppendGraphicsExtensions{.gif}

\bibpunct{(}{)}{;}{a}{}{,}

\begin{document}
\title{A transiting companion to the eclipsing binary KIC002856960}
\author{D. Armstrong$^1$  \and D. Pollacco$^1$ \and C. A. Watson$^2$ \and F. Faedi$^1$ \and Y. G\'{o}mez Maqueo Chew$^{1,4}$ \and H. M. Cegla$^{2,4}$ \and P. McDaid$^2$ \and J. Burton$^2$ \and J. McCormac$^3$ \and I. Skillen$^3$}

\institute{Department of Physics, University of Warwick, Gibbet Hill Road, Coventry CV4 7AL, UK \email{darmstrong25@qub.ac.uk}  \and Astrophysics Research Centre, School of Mathematics \& Physics, Queen's University Belfast, University Road, Belfast BT7 1NN, UK \and Isaac Newton Group of Telescopes, Apartado de Correos 321, E-38700 Santa Cruz de la Palma, Canary Islands, Spain \and Department of Physics \& Astronomy, Vanderbilt University, Nashville, TN 37235, USA}

\date{Received 26 April 2012 / Accepted 17 August 2012 }
\abstract{
We present an early result from an automated search of \emph{Kepler} eclipsing binary systems for circumbinary companions. An intriguing tertiary signal has been discovered in the short period eclipsing binary \object{KIC002856960}. This third body leads to transit-like features in the light curve occurring every 204.2 days, while the two other components of the system display eclipses on a 6.2 hour period. The variations due to the tertiary body last for a duration of {\raise.17ex\hbox{$\scriptstyle\sim$}}1.26 days, or 4.9 binary orbital periods. During each crossing of the binary orbit with the tertiary body, multiple individual transits are observed as the close binary stars repeatedly move in and out of alignment with the tertiary object. We are at this stage unable to distinguish between a planetary companion to a close eclipsing binary, or a hierarchical triply eclipsing system of three stars. Both possibilities are explored, and the light curves presented.
}

\maketitle

\section{Introduction}
With approximately a third of main sequence stars in the galaxy found in binary systems \citep{Lada:2006kp} but a much smaller percentage of known planets in such, planets in binary systems represent an important area of investigation. Transiting circumbinary planets can provide a wealth of information about themselves and their host systems, from the extreme system flatness necessary for planetary transits and binary eclipses to be seen to potential absolute dynamical solutions once timing variations are taken into account. As such they are primary targets for attempted detection. The \emph{Kepler} mission has already discovered three circumbinary planets \citep{2011arXiv1109.3432D,Welsh:2012kl} and provides the opportunity for many more. The components of these three systems are all well separated, leaving the parameter space of transiting planets around contact and near contact binaries as yet unexplored.

Triple stellar systems are also of great interest, and several have recently been detected through the \emph{Kepler} data \citep{Derekas:2011js,Steffen:2011ie}. If such systems transit, their signals can be similar to those from planetary systems, and so a search for planets in binaries may also lead to detection of triple stars. These can provide insights into the stability, dynamics and evolution of complex systems. Recently such a triply eclipsing triple was discovered and characterised using data from the \emph{Kepler} mission \citep{Carter:2011kx}, providing rare observational constraints for low mass stars with fully convective interiors.

With the aim of expanding the parameter space of known circumbinary planets a search was performed on the eclipsing binaries listed by the Kepler Eclipsing Binary Catalogue (KEBC) \citep{Prsa:2011dx,Slawson:2011fg}, initially on those systems not classified as detached. Here we present one early discovery from the search: \object{KIC002856960} (RA(J2000) 19:29:31.52, DEC +38:04:35.9, \emph{Kepler} magnitude 15.615), a system composed of either a close eclipsing binary orbited by a transiting planetary companion or low mass dwarf star, or a triply eclipsing group of at least three low mass stars. This result and others in the future should prove of value to models of complex systems, where gravitational interactions of several stars can lead to unusual perturbations (e.g. \citet{QUINTANA:2006dy} in the planetary case), and as such provide fundamental observational information to general formation models.

\section{Observations}

\subsection{Data}
Observations of the KEBC systems were taken by the NASA \emph{Kepler} satellite, a mission producing extremely high precision, near continuous light curves of {\raise.17ex\hbox{$\scriptstyle\sim$}}155,000 stars on the level of 20 ppm \citep{2010ApJ...713L..79K,2010ApJ...713L.109B,2010ApJ...713L.115H}. This mission began science operations  on 13 May 2009, and data up to the end of Quarter 6 (22 Sep 2010) are publicly available on the NASA Data Archive\footnote{http://archive.stsci.edu/kepler/}. Due to technical constraints, the satellite must reorient itself each quarter year, and so data is provided as a separate file per object and per quarter, each quarter consisting of three months of observations. Supplementary information on targets and their designations are available from the Kepler Input Catalogue (KIC) \citep{Brown:2011dr}. The data presented were taken in long cadence mode (29.4 minute exposures) over a period of 498 days, and represent 22651 good exposures when small gaps in the data are taken into account. See the Data Characteristics Handbook on the NASA Data Archive\footnote{http://archive.stsci.edu/kepler/documents.html} for precise observation periods. In raw format, the data show a variety of noise signals, the details of which are explained in the Kepler Data Processing Handbook\footnotemark[\value{footnote}]. The removal of the majority of the instrumental parts of these was performed using covariance basis vectors (CBVs), enacted using the PyKe python module\footnote{http://keplergo.arc.nasa.gov/PyKE.shtml}. This method attempts to remove instrumental effects by using variations observed in several stars neighbouring the object of interest, and as such should robustly maintain astrophysical signals. For a full explanation of the method see the software description\footnote{http://keplergo.arc.nasa.gov/ContributedSoftwareKepcotrend.shtml}.

\subsection{Eclipsing Binary Signal Removal}
\label{ebsigrem}
Data were then processed to remove the binary light signature from each quarter separately (hereafter `whitening'). First, data were phase folded using the period provided by the KEBC. The phased curve was then binned using 200 equal width bins, and the median of each bin determined. As any points exhibiting tertiary transit events will be distributed across the light curve when phase folded on the binary period, taking the median will exclude them from the whitening process. The median of each bin is then subtracted from each point in the bin, resulting in a series of data points distributed around a median adjusted flux of zero. This method has the advantage of requiring no knowledge of the signal being removed beyond its rate of variance (required to allow the bin number to be selected appropriately), and as such can be applied to many variable objects. The whitened quarters were then concatenated and searched for tertiary events. An automated search method (to be described fully in Armstrong et al. 2012b,in prep) combined with by-eye inspection of the results led to the discovery of the object presented here.

\begin{figure}
  \resizebox{\hsize}{!}{\includegraphics{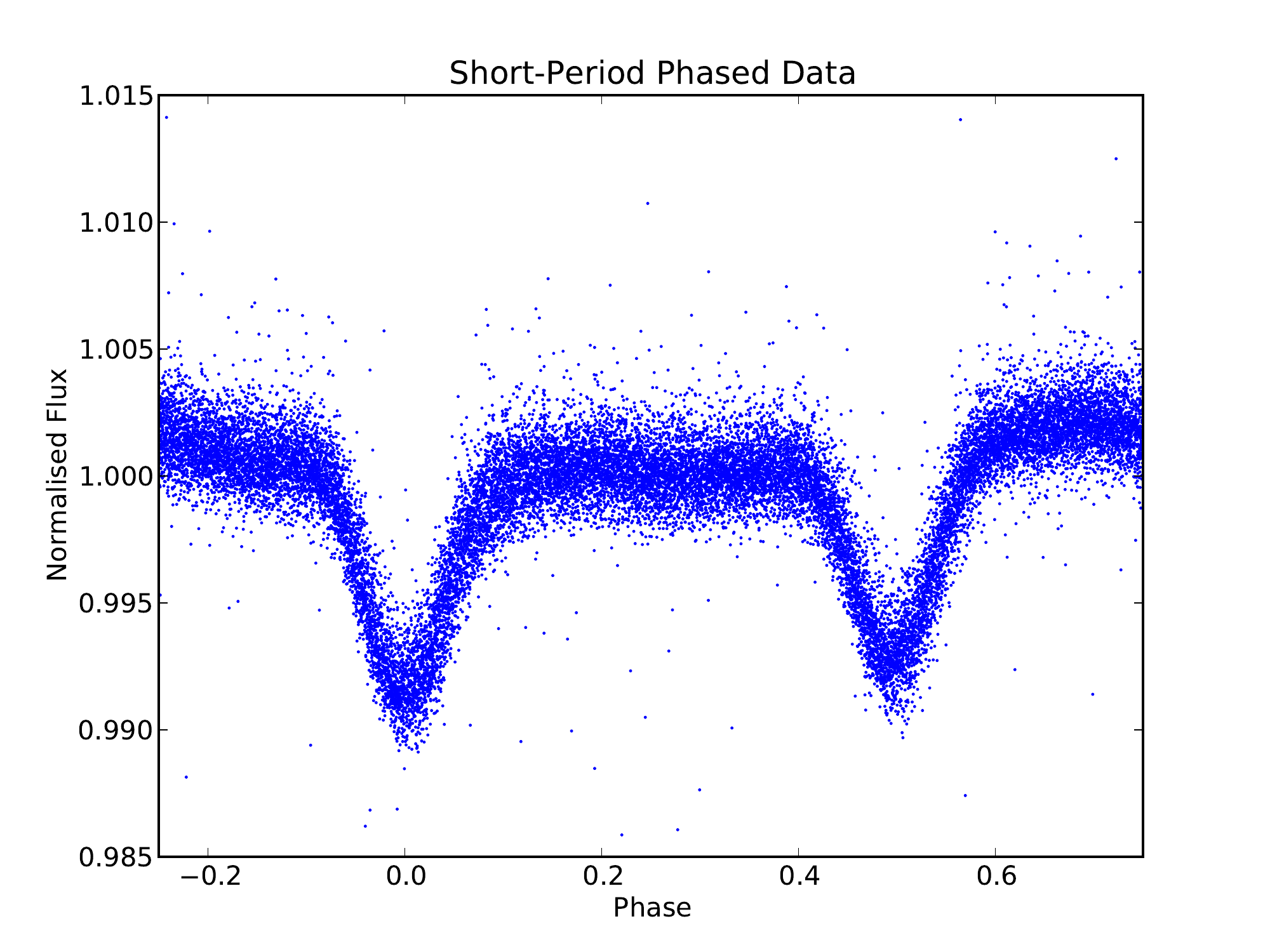}}
  \caption{The binary variation (eclipses of A and B by each other), shown as phase folded data from all 6 quarters (scale shown is relative offset to the median of each quarter). The tertiary event has not been removed, and as such some points are omitted below the figure window. These are displayed more clearly in Fig. \ref{tertlc}. Error bars are not plotted, as the spread of points gives a much clearer indication of the true error.}
  \label{binlc}
\end{figure}

\section{Results}

\begin{figure*}
\centering
 \includegraphics[width=17cm]{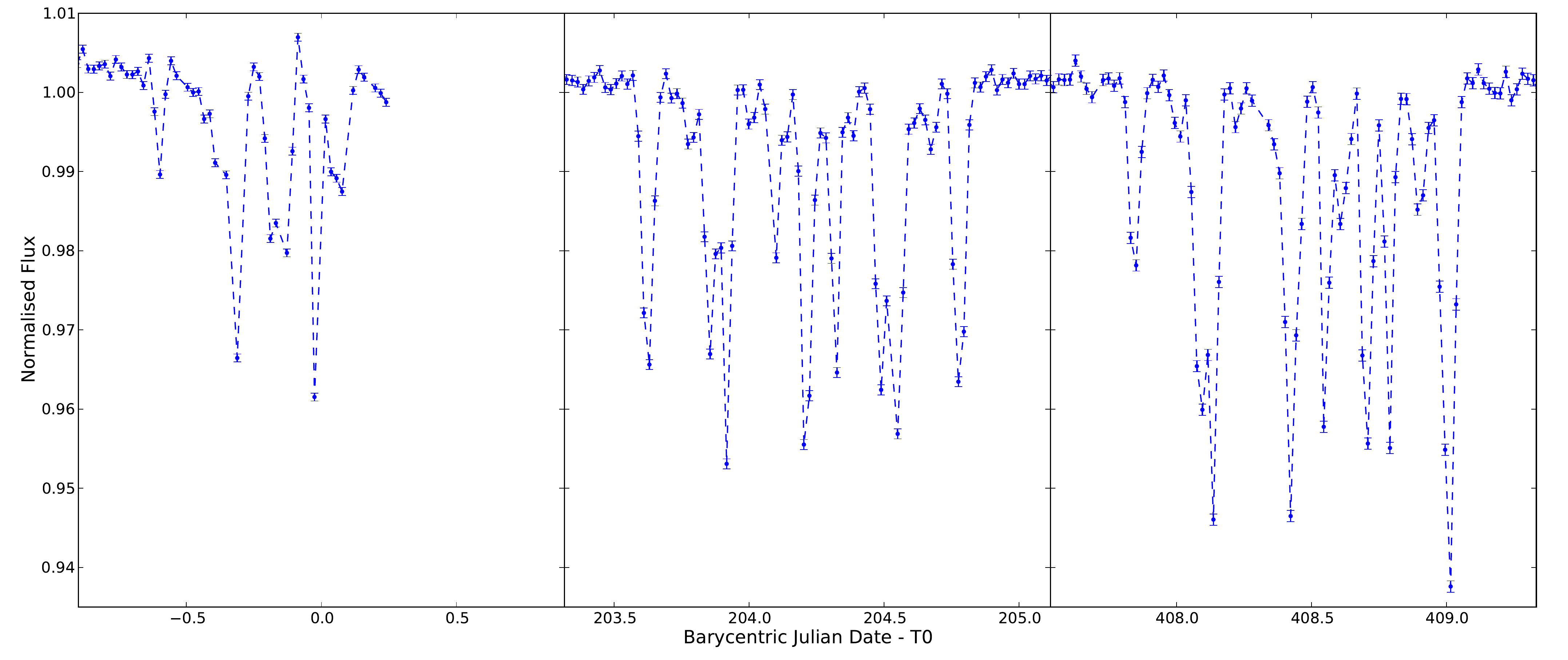}
  \caption{The three tertiary crossings available in the public data, in quarters 1, 4 and 6. Crossings are referred to as Crossing 1, 2 and 3, from left to right respectively. Light curves are displayed after removal of the binary signal. Dashed lines are simple connections of data points, for clarity, and do not stem from a model.}
  \label{tertlc}
\end{figure*}

\subsection{Overview}
Periodic signals separate from the previously known short period variations shown in Fig. \ref{binlc} have been detected from the object \object{KIC002856960}. There are three new events associated with the long period signal seen in the public data, in quarters 1, 4 and 6, and these are separated and presented in Fig. \ref{tertlc}. They display a unique appearance, indicative of the presence of at least three bodies in the system.\\
\indent The short period variations have not been amenable to modelling by the catalogue team. As such, while we proceed under the assumption that the source of the short period variation is an eclipsing binary system (of a perhaps complex nature), no binary parameters beyond the ephemeris of the short-period variation are provided. We have derived an updated binary orbital period for this signal using all of the publicly available quarters (1-6) via $\chiup^2$ minimisation of each bin in the binned data (see Sect. \ref{ebsigrem}) and present this along with the catalogue value for $\mathrm{T}_{0}$ in Table \ref{resultstable}. The original binary signal is approximately 1\% in depth, notably shallower than the newly detected signal.\\
\indent From here we shall refer to the bodies in this system as Components AB (for the two members of the binary system), and either b or C depending on whether the third body is considered as a planet or star respectively. We limit ourselves to considerations of three bodies only, for simplicity.\\
\indent Each primary new event (hereafter `crossing', as each event consists of a crossing with the binary orbit by a third body) contains multiple transits of A or B before C, or b/C before A or B (the distinction depends on the system architecture, see Sect. \ref{Disc}). Ephemerides for the three observed crossings are given, along with durations and depths, calculated from the deepest point observed in each crossing. The implications of these results for our interpretation of the nature of the system are discussed in Sect. \ref{Disc}. The crossing duration is given for comparison with the binary period - the first crossing in the quarter 1 light curve is cut off by the end of the quarter before completion of the crossing, and hence a duration is not provided.\\
\indent If secondary eclipse events in the long period signal (of component b/C by AB, or AB by C) are present, they are constrained by the variability of the data to have an upper limit to their depth of {\raise.17ex\hbox{$\scriptstyle\sim$}}0.4\%, and no evidence for any is observed in the phased light curve.\\
\indent Time based parameter errors are dominated by the long cadence, in the absence of a full model. Crossings are taken as beginning midway between the data point immediately before flux variations are seen and its successor (the choice of these points is clear on viewing the data), and the absolute errors set such that the beginning of ingress could happen anywhere between these points. End times are derived similarly. All parameter errors derive from formal propagation from these values or the raw flux errors as appropriate.

\subsection{Blending}
\label{blending}

\begin{figure}
  \centering
  \includegraphics[scale=0.6]{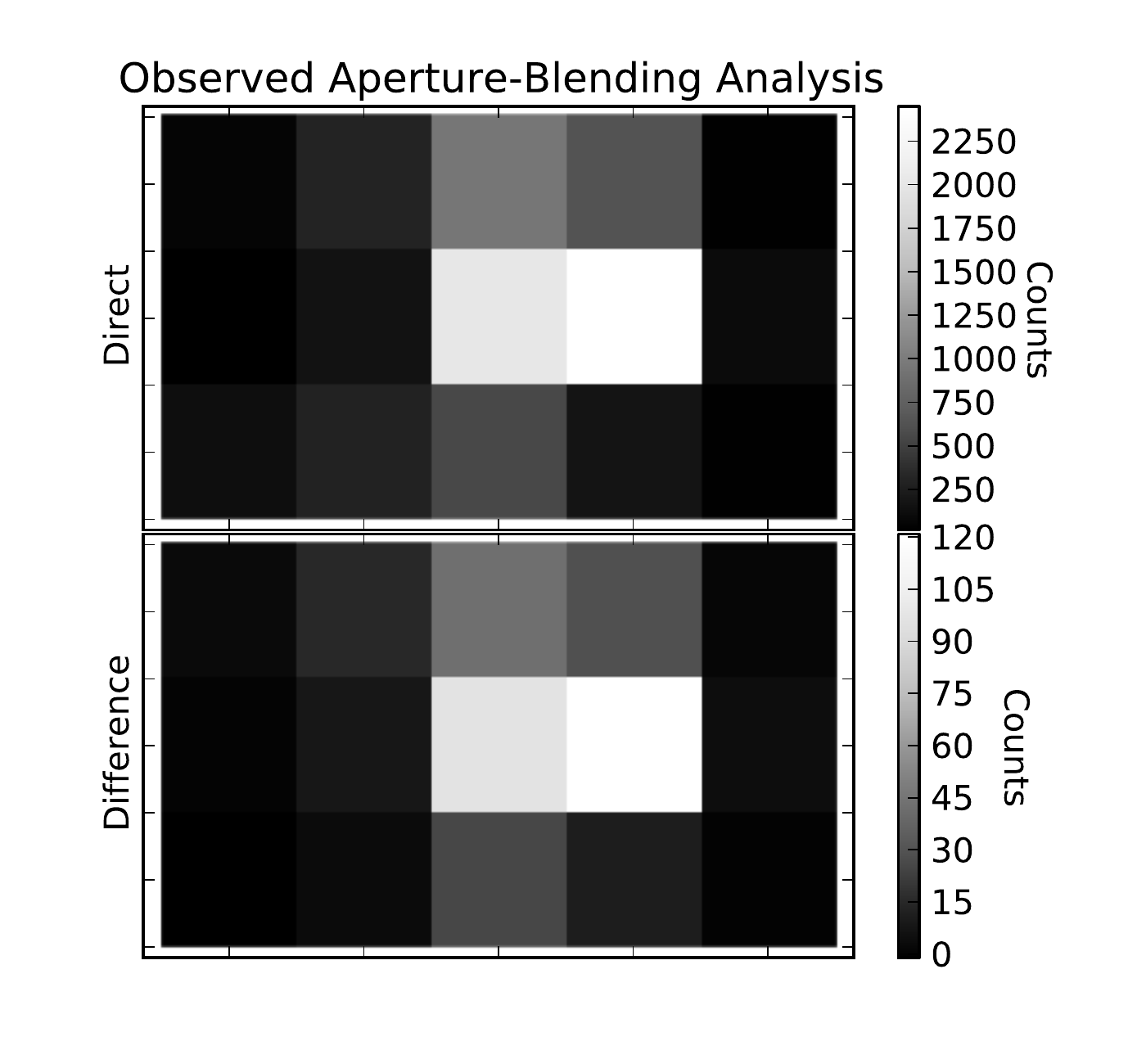}
  \caption{Images formed from the observed Q4 aperture for \object{KIC002856960}, using the method as utilised by \citet{Torres:2010eo}. The direct image shows the location of the primary source, whereas the difference image shows the location of the flux variation. No discernible difference is observed.}
  \label{centroidfig}
\end{figure}

It has been noted that the depths for each transit are markedly different; this is expected but quarter to quarter variations in the depth, along with the large size of the \emph{Kepler} pixels of  {\raise.17ex\hbox{$\scriptstyle\sim$}}4 arc sec \citep{2010ApJ...713L..79K} can cause suspicion as to whether the object involved is blending with a background source. The photometric aperture used changes between quarters to optimise the photometric precision, meaning that a significant background source may contribute a different amount of flux each quarter. \object{KIC002856960} has been subjected to a centroid analysis as utilised by \citet{Torres:2010eo}. This involves creating a difference image across the pixel aperture between a selection of points in and out of transit. This difference image then shows the `true' location of the flux variation. This is compared to a direct image formed from summing the same points. The direct image shows the location of the primary source of flux. Fig. \ref{centroidfig} shows the difference image produced, and the comparison direct image. Blends originating from an offset of order 1 pixel are easily ruled out by inspection. The fit of a pixel response function (a modified point spread function - \citet{Bryson:2010ci}) limited the region where a blending source could be located to within a radius of order 0.3 pixels from the target, but was complicated by the presence of the binary signal. Despite the difficulties this does rule out a large region of the photometric aperture as the location of a background varying source.\\
\indent The nature of the tertiary signal requires that three bodies are present in the system; for it to arise from a separate system to the binary flux variations, while being so closely positioned, is extremely unlikely. The depth of the tertiary signal also indicates that, if it was caused by a background source, the contaminating source would need to be of very significant brightness compared to the target. No evidence for such a source is seen. As such, we find it very likely that \object{KIC002856960} is indeed the source of the signals presented in this paper.

\begin{table}
\caption{Derived parameters for the \object{KIC002856960} system}           
\label{resultstable} 
\begin{tabular}{lccc}
\hline\hline
Parameter &Units & Result & Error (absolute) \\    
\hline
Period (AB) &d& 0.2585082 & ... \\
$\mathrm{T}_0$ (AB)&BJD&54964.652115&...  \\
Period (b/C) &d& 204.2163 & 0.030\\
$\mathrm{T}_0$ (b/C)&BJD&  54997.743313&0.024667 \\
Duration (b/C)&d& & \\
\hspace{2em}Crossing 2 &&1.2464&0.0204\\
\hspace{2em}Crossing 3 &&1.2669&0.0204\\
Depth (b/C)&\%& &\\
\hspace{2em}Crossing 1&&4.1171&0.0588\\
\hspace{2em}Crossing 2 &&4.8179& 0.0669\\
\hspace{2em}Crossing 3 &&6.3207&0.0749\\
\hline
\end{tabular}
\tablefoot{Parameters are labelled AB or b/C, standing for signals associated with the binary stars (AB) or companion (b/C). BJD stands for Barycentric Julian Date. $\mathrm{T}_{0}$ given as $\mathrm{T}_{0}$ - 2400000. A crossing 1 duration is not provided as this crossing is cut off by the end of a quarter before completion.}
\end{table}

\section{Discussion}
\label{Disc}
\subsection{Summary}
Our primary interest lies in the tertiary signal detected. The public data contains three crossings (Fig. \ref{tertlc}) appearing in quarters 1, 4 and 6. These crossings occur on a period much longer than that of the binary AB, and are indicative of the presence of a companion third body b/C either crossing or being crossed by the binary AB. The short period binary completes {\raise.17ex\hbox{$\scriptstyle\sim$}}5 orbits during the course of the crossing, meaning that during a crossing multiple transits will be produced as the components A or B move in and out of alignment with the slower moving third body b/C. The shape of the transits seen within a crossing will be dependent on the velocity and relative positions of all components, as well as the components' shape and surface brightness. In the absence of a full model or complete mapping of the flux variations (due to the long cadence data only just resolving the inner transits within each crossing) we have presented the simplest observable parameters available: ephemeris, depth and overall crossing duration. Since the ingress/egress times are expected to depend on the location of all three bodies at the start/end of each tertiary event, an accurate period cannot be determined with only three crossing events. This situation will be largely improved once the currently private data becomes accessible.\\
\indent The structure and nature of this system remain in doubt. Here we discuss the two most likely system geometries: an object in orbit around a close binary pair (system (AB)b or (AB)C), or a close binary pair in orbit around a larger star (system C(AB), where C is the more massive star in this case). We consider these two options in turn, acknowledging that there may be other more unusual configurations possible. The equal depths of the AB binary eclipses (Fig. \ref{binlc}) imply that, if the binary components AB were on the main sequence, they should be of roughly equal mass. Given their short period, we find it plausible that the binary components may have evolved away from the main sequence, although without further data this cannot be constrained. In either case, the equal depth of the AB binary eclipses shows that the components A and B must have the same surface brightness.

\subsection{Circumbinary Object}
\label{circumbin}

In this scenario the tertiary body b/C is of lower mass than the combined binary pair AB. A secondary eclipse of b/C by A or B that is not observable implies that the majority of the flux must be emitted from the components AB.  Here we use the KIC colour derived parameters to explore the nature of this complex system.\\
\indent The KIC temperature (4733K) and stellar radius imply a star of luminosity $0.26L_{\odot}$. If both central stars AB were of this type, the 0.4\% limit on secondary eclipses limits the luminosity of b/C to be {\raise.17ex\hbox{$\scriptstyle\sim$}}$0.002L_{\odot}$ in this scenario. This is consistent with an object of spectral type M5 or cooler \citep{reidhawley}. Presenting a specific companion radius is beyond our capability at this stage, but below we present example companion radii leading this configuration to support a circumbinary dwarf star, brown dwarf or perhaps transiting planet around a close binary pair, in the configuration (AB)b/C. The extremely shallow AB eclipses evident in the binary lightcurve could possibly be explained as the result of a grazing geometry.\\
\indent The deepest observed transit (6.32\%), on a single star of KIC stellar radius $0.757R_\odot$ \citep{Brown:2011dr}, implies a companion radius of $1.85R_J$, or $0.19R_\odot$. Assuming AB components of equal radii (and equal surface brightness as shown by the depth of their mutual eclipses), dilution would increase the companion radius to $0.27R_\odot$. Given the very short period of AB it is possible that the colour derived stellar radius will be an overestimate, allowing for the companion radius to decrease. There remains the possibility of the observed eclipses of A or B by component b/C being grazing, in which case the `true' transit depth would be greater and the companion radius also.

\subsection{Circumstellar Binary}
This scenario assumes that the tertiary body is of greater mass than the binary pair. As such it would be the source of the majority of the flux, and can be taken to be the object described by the KIC (consistent with a K star). This would then be orbited by the lower mass stellar binary pair AB, in the architecture C(AB).  Taking the KIC stellar radius of the primary K star component C, we again derive a radius for a single component of the binary system A or B of $0.19R_{\odot}$. The system would then consist of a central K star orbited by a close pair of remarkably small dwarf stars. This has the advantage of explaining the shallow AB binary eclipse depths observed through dilution by the K star. The lack of secondary eclipses (eclipses of A or B by component C) imposes a problem however. To produce secondary eclipses of order 0.4\% or less, such that they would not be visible in the data as shown, requires that each component of the binary be contributing less than 0.4\% of the system flux (for total eclipses). This is not possible given the {\raise.17ex\hbox{$\scriptstyle\sim$}}1\% binary eclipse depths observed (where the components would each need to be contributing at least 1\% of the system flux). As such this configuration requires that the secondary eclipses be grazing, and thereby the primary eclipses also. Without further modelling we cannot take either configuration further.
\subsection{Conclusion}
At this stage we are unable to conclusively distinguish between the two system architectures (AB)b/C or C(AB) as outlined above. We also acknowledge that there may well be other stable system configurations unexplored here. Spectral follow up and the currently private \emph{Kepler} data, in combination with the presented lightcurves, will allow for a much greater understanding of this intriguing system. When its nature is fully understood the absolute determination of many of the system parameters will be possible. This includes the companion mass in the (AB)b/C case, if transit timing variations as used in \cite{2011arXiv1109.3432D} and \cite{Welsh:2012kl} prove significant. We hope that this will prompt follow up studies, and will continue to develop our understanding of the object as more data becomes available. If the (AB)b case proves true, this system represents the first discovered transiting planet around such a relatively short period eclipsing binary. If the C(AB) or (AB)C cases prove true this system will contain stars with some of the lowest masses accurately determined. These will be low enough to imply fully convective interiors, an area where directly measured physical properties are few (e.g. \citep{Carter:2011kx,Morales:2009dc,Irwin:2011fo}).

\begin{acknowledgements}
We thank the anonymous referee for making comments which significantly improved the paper. This paper includes data collected by the \emph{Kepler} mission. Funding for the \emph{Kepler} mission is provided by the NASA Science Mission directorate. All of the data presented in this paper were obtained from the Mikulski Archive for Space Telescopes (MAST). STScI is operated by the Association of Universities for Research in Astronomy, Inc., under NASA contract NAS5-26555. Support for MAST for non-HST data is provided by the NASA Office of Space Science via grant NNX09AF08G and by other grants and contracts. DA, PM and JB are supported by DEL grants. HMC acknowledges support from a QueenÕs University Belfast university scholarship. CAW would like to acknowledge support by STFC grant ST/I001123/1.
\end{acknowledgements}

\end{document}